\documentclass[twocolumn,showpacs,preprintnumbers,amsmath,amssymb,prl]{revtex4}

\usepackage{graphicx}
\usepackage{dcolumn}
\usepackage{bm}
\usepackage{color}
\usepackage{soul}
\usepackage{notes2bib}

\begin{document}



\title{Observation of the universal magnetoelectric effect in a 3D topological insulator}

\author{V.~Dziom$^{1}$}
\author{A.~Shuvaev$^{1}$}
\author{A.~Pimenov$^{1}$}
\email[E-mail:~]{pimenov@ifp.tuwien.ac.at}
\author{G.~V.~Astakhov$^{2}$}
\email[E-mail:~]{astakhov@physik.uni-wuerzburg.de}
\author{C.~Ames$^{4}$}
\author{K.~Bendias$^{4}$}
\author{J.~B\"ottcher$^{3}$}
\author{G.~Tkachov$^{3}$}
\author{E.~M.~Hankiewicz$^{3}$}
\author{C. ~Br\"{u}ne$^{4}$}
\author{H~Buhmann$^{4}$}
\author{L.~W.~Molenkamp$^{4}$}

\affiliation{$^{1}$Institute of Solid State Physics, Vienna University of Technology, 1040 Vienna, Austria \\
$^{2}$ Physikalisches Institut (EP6), Universit\"{a}t W\"{u}rzburg, 97074 W\"{u}rzburg, Germany \\
$^{3}$ Institut f\"{u}r Theoretische Physik und Astronomie, Universit\"{a}t W\"{u}rzburg, 97074 W\"{u}rzburg, Germany \\
$^{4}$ Physikalisches Institut (EP3), Universit\"{a}t W\"{u}rzburg, 97074 W\"{u}rzburg, Germany}

\begin{abstract}
The electrodynamics of topological insulators (TIs) is described by modified Maxwell's equations, which contain additional terms that couple an electric field to a magnetization and a magnetic field to a polarization of the medium, such that the coupling coefficient is quantized  in odd multiples of $e^2 / 2 h c $ per surface. Here, we report on the observation of this so-called topological magnetoelectric (TME) effect. We use monochromatic terahertz (THz) spectroscopy of TI structures equipped with a semi-transparent gate to selectively address surface states. 
In high external magnetic fields, we observe  a universal Faraday rotation angle equal to the fine structure constant $\alpha = e^2 / \hbar c$ when a linearly polarized THz radiation of a certain frequency passes through the two surfaces of a strained HgTe 3D TI.  These experiments give insight into axion electrodynamics of TIs and may potentially be used  for a metrological definition of the three basic physical constants. 
 
\end{abstract}

\date{\today}

\maketitle

Maxwell's equations  are in the foundation of modern optical and electrical technologies. In oder to apply Maxwell's equations in conventional matter, it is necessary to specify constituent  relations, describing the polarization $ \mathbf{P}_c (\mathbf{E})$ and magnetization $\mathbf{M}_c (\mathbf{B})$ as a function of the applied electric and magnetic fields, respectively. Soon after the theoretical prediction \cite{PhysRevLett.95.226801, Bernevig15122006, PhysRevB.76.045302} and experimental discovery of 2D and 3D TIs \cite{Koenig02112007, Hsieh2008}, it has been recognized that the constituent  relations in this new phase of quantum matter contain additional cross-terms $ \mathbf{P}_t (\mathbf{B})$  and $ \mathbf{M}_t (\mathbf{E})$  \cite{PhysRevB.78.195424}
\begin{equation}
 \begin{split}
 & \mathbf{P}_t  ( \mathbf{B}) = \left( N + \frac{1}{2} \right) \frac{\alpha}{2 \pi} \mathbf{B} \, \\
 & \mathbf{M}_t (\mathbf{E}) = - \left( N + \frac{1}{2} \right) \frac{\alpha}{2 \pi} \mathbf{E} \,.
 \end{split}
 \label{Maxwell}
\end{equation}
Here, $N$ is an integer, and $\alpha \approx 1 / 137$ is the fine structure constant. Intriguing consequences of Eq.~(\ref{Maxwell}) are the universal Faraday rotation angle $| \theta_F | = \alpha$, when a linearly polarized electromagnetic radiation passes through the top and bottom topological surfaces \cite{PhysRevB.78.195424, PhysRevLett.105.057401}, and magnetic monopole images, induced by electrical charges in proximity to a topological surface \cite{Qi27022009}. However, experimental verification of these TME effects has been lacking. 

As the modified Maxwell's equations describing electrodynamics of TIs are applicable in the low-energy limit, optical experiments should be performed at THz or sub-THz frequencies \cite{PhysRevLett.107.136803, PhysRevLett.108.087403, PhysRevB.87.121104, Wu2013}. In real samples, the TME may be screened by nontopological contributions  \cite{PhysRevLett.105.166803, PhysRevB.82.161104, PhysRevB.84.035405}. In fact, quantized Faraday rotation has been detected in 2D electron gas \cite{PhysRevLett.104.256802} and graphene \cite{Shimano2013} in the quantum Hall effect (QHE) regime, but the observed values are not fundamental. 

We report on the observation of the universal Faraday rotation angle equal to the fine structure constant $\alpha$. Strained HgTe layers grown on CdTe, that are investigated in the present work, are shown to be a 3D TI \cite{PhysRevLett.106.126803} with surface-dominated charge transport \cite{PhysRevX.4.041045}.  In oder to eliminate the material details, we perform measurements under constructive interference conditions, such that the transmission through the CdTe substrate is approaching 100\% \cite{PhysRevLett.106.107404}. 

\begin{figure*}[t]
\includegraphics[width=.77\textwidth]{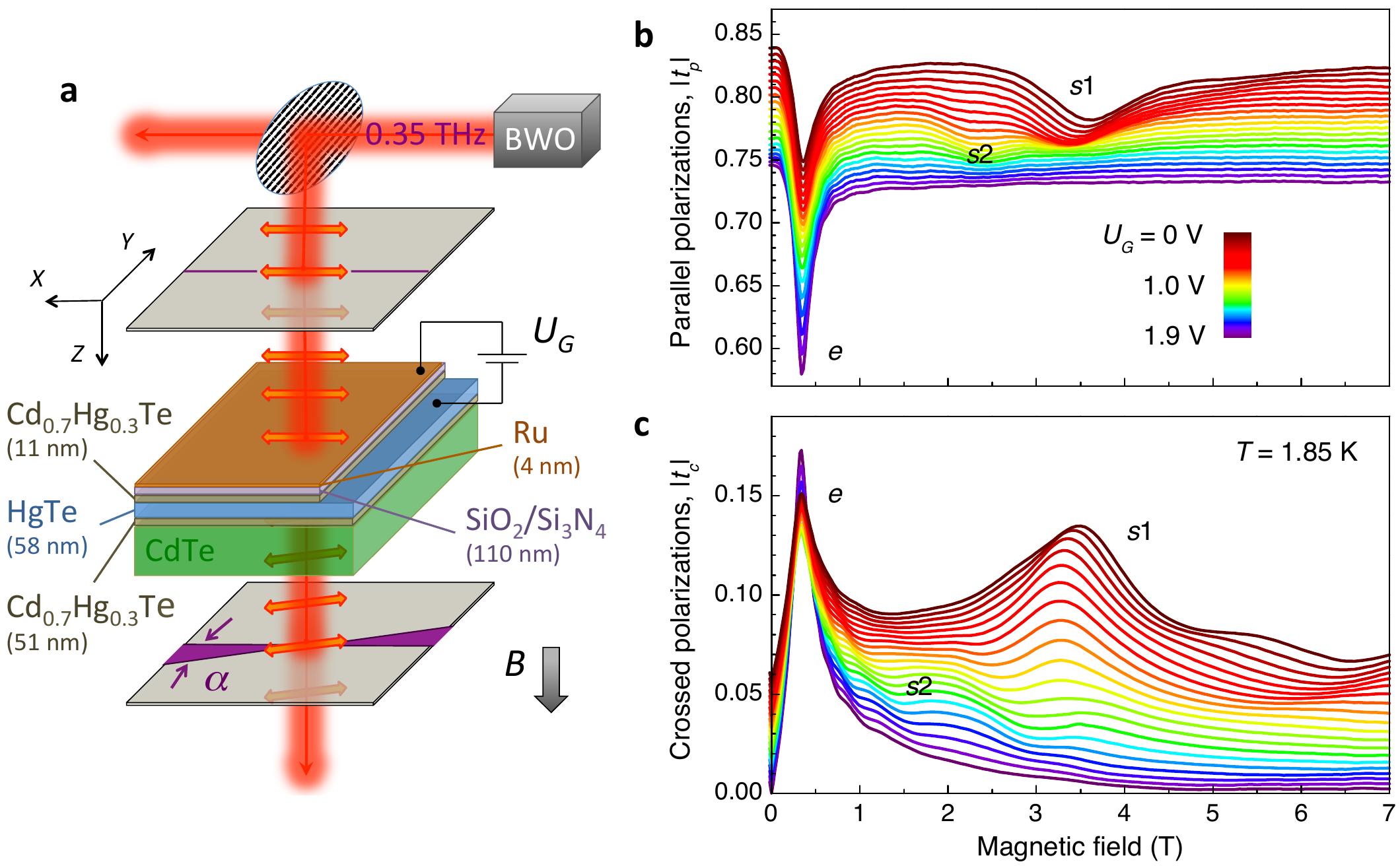}
\caption{\textbf{THz magnetooptics of a strained HgTe 3D TI.} \textbf{a},  A scheme of the experimental setup (only one arm of the Mach-Zehnder interferometer is shown).  The strained HgTe layer, which is a 3D TI, is sandwiched between (Cd,Hg)Te protecting layers. The top-gate electrode, consisting of a SiO$_2$/Si$_3$N$_4$ multilayer insulator and a thin conducting Ru film, is semitransparent at THz frequencies. The THz radiation ($\nu = 0.35 \, \mathrm{THz}$) is linearly polarized, and the Faraday rotation ($\theta_F$) and ellipticity ($\eta_F$) are measured as a function of the magnetic field $B$ for different gate voltages $U_G$. \textbf{b}, \textbf{c}, Transmission spectra in the parallel $| t_p |$ and crossed $| t_c |$ polarizer configurations, respectively. The gate voltage is color-coded, and the experimental curves are shifted for clarity. Notations in \textbf{b} and \textbf{c}: $e$ denotes the CR of the topological surface states of electron character,   $s1$ and $s2$ denote extra resonances with opposite phase to that of the $e$-CR as discussed in the text. } \label{fig1}
\end{figure*}

The strained HgTe film is a 58 nm thick HgTe layer embedded between two Cd$_{0.7}$Hg$_{0.3}$Te layers (Fig.~\ref{fig1}a). The Cd$_{0.7}$Hg$_{0.3}$Te layers have a thickness of  $51 \, \mathrm{nm}$ (lower layer) and $11 \, \mathrm{nm}$  (top/cap layer), respectively. The purpose of these layers is to provide the identical crystalline interface for top and bottom surface of the HgTe films as well as to protect the HgTe from oxidization and adsorption. This leads to an increase in carrier mobility with a simultaneous decrease in carrier density compared to uncaped samples \cite{PhysRevLett.106.126803}. The transport characterization on a standard Hall bar sample shows a carrier density at $0 \, \mathrm{V}$ gate of $1.7 \times 10^{11} \, \mathrm{cm^{-2}}$ and a carrier mobility of $2.2 \times 10^{5} \, \mathrm{cm^{2} V^{-1} s^{-1}}$. The optical measurements are carried out on a sample fitted with a $110 \, \mathrm{nm}$ thick multilayer insulator of SiO$_2$/Si$_3$N$_4$ and a $4 \, \mathrm{nm}$ thick Ru film. The Ru film (oxidized in the air) is used as a semitransparent top gate electrode \cite{APL4811496}.

The transmittance experiments at THz frequencies ($0.1 \, \mathrm{THz} <  \nu  < 1 \, \mathrm{THz}$) are carried out in a Mach-Zehnder interferometer arrangement \cite{volkov_infrared_1985, shuvaev_sst_2012}, allowing  measurement of the amplitude and the phase shift of the electromagnetic radiation in a geometry with controlled polarization (Fig.~\ref{fig1}a). The monochromatic THz radiation is provided by a backward-wave oscillator (BWO). The THz power on the sample is in between $10  \, \mathrm{\mu W}$  and $100  \, \mathrm{\mu W}$  with the focal spot of $0.2  \, \mathrm{cm^2}$. Using wire grid polarizers, the complex transmission coefficient $t = | t | e^{i \phi}$ is obtained both in parallel $t_p $ (Fig.~\ref{fig1}b) and cross $t_c $ (Fig.~\ref{fig1}c) polarization geometries, providing full information about the transmitted light.  External magnetic fields $B \leqslant 7 \, \mathrm{T}$ are applied  using a split-coil superconducting magnet. The experiments are carried out in Faraday geometry, i.e with $B$ applied parallel to the propagation direction of the THz radiation. The $ac$ conductivity tensor $\hat{\sigma} (\omega)$  at THz angular 
frequency $\omega = 2 \pi \nu$ is obtained from the experimental data by inverting the Berreman equations \cite{berreman_josa_1972} for the complex transmission coefficient through a thin conducting film on an insulating substrate. 

In general case, the light propagating along the $z$ direction can be characterized by the orthogonal $x$ and $y$ components of the electric  and magnetic  fields, which can be written in the form of a 4D vector $\mathbf{V}$. The interconnection between vectors $\mathbf{V}_1$ and $\mathbf{V}_2$, corresponding to different points in space separated by a distance $\ell$, is given by $\mathbf{V}_1 = \hat{M}(\ell) \mathbf{V}_2$. Here, $\hat{M}(\ell)$ is a $4 \times 4$ transfer matrix. For an insulating substrate of thickness $\ell$ and dielectric constant $\varepsilon$, this is the identity matrix $\hat{M}_{\mathrm{CdTe}}(\ell) = \mathbb{I}$ provided $\ell \sqrt{\varepsilon} \, \nu / c$ is an integer. We find in a separate experiment on a bare CdTe substrate that this condition is fulfilled for $\nu \approx 0.35 \, \mathrm{THz}$. Therefore, all the measurements presented here are performed at this frequency to eliminate any contribution to the Faraday signal from the substrate. The corresponding photon energy of $1.4 \, \mathrm{meV}$ is much smaller than the energy gap in strained HgTe (above $10 \, \mathrm{meV}$) \cite{PhysRevLett.106.126803},  and Eqs.~(\ref{Maxwell}) are a good approximation. 

For normal incidence, the fields across the conducting interface are connected by the Maxwell equation $\nabla \times
\mathbf{H} = \hat{\sigma} \mathbf{E}$. Here, the $e^{-i\omega t}$ time dependence is assumed for all fields.  As the wavelength of $856 \, \mathrm{\mu m}$ for $\nu = 0.35 \, \mathrm{THz}$ is much larger than the HgTe layer thickness, we use the limit of thin film, and the corresponding transfer matrix $\hat{M}_{\mathrm{HgTe}}(\hat{\sigma})$ is determined by the diagonal ($\sigma_{xx} $) and Hall ($\sigma_{xy} $) components of the conductivity tensor $\hat{\sigma}$. Within the Drude-like model, these components for one type of charge carriers can be written in the form \cite{palik_rpp_1970, PhysRevB.82.161104} 
\begin{eqnarray}
&& \sigma_{xx} =\sigma_{yy}  = \frac{1-i \omega \tau}{(1-i \omega
\tau)^2 +(\Omega_c \tau)^2} \sigma_0
\,, \label{sxx}\\
&& \sigma_{xy} =-\sigma_{yx} = \frac{\Omega_c \tau}{(1-i \omega
\tau)^2 +(\Omega_c \tau)^2} \sigma_0 \,. \label{sxy}
\end{eqnarray}
Here, $\Omega_c$ is the cyclotron resonance (CR) frequency, $\sigma_0$ is the $dc$ conductivity, and $\tau$ is the scattering time. For classical conductors, the CR frequency is written as $\Omega_c = eB/m_{e}$, where $m_{e}$ is the effective electron mass in the parabolic approximation. 

The total transfer matrix $\hat{M} = \hat{M}_{\mathrm{CdTe}}  \hat{M}_{\mathrm{HgTe}}$ relates vectors $\mathbf{V}$ on both sides of the sample and hence contains full information about the transmission and reflection coefficients.  Thus, when $\hat{M}_{\mathrm{CdTe}} = \mathbb{I}$, the influence of the substrate is minimized, and the THz response  is dominated by the $ac$ transport properties of the HgTe layer, in accord with Eqs.~(\ref{sxx}) and (\ref{sxy}).  The calculation of the complex transmission coefficients $t_p$ and $t_c$ based on the transfer matrix formalism as well as the exact form of the transfer matrices are presented in Methods.  

\begin{figure}[t]
\includegraphics[width=.43\textwidth]{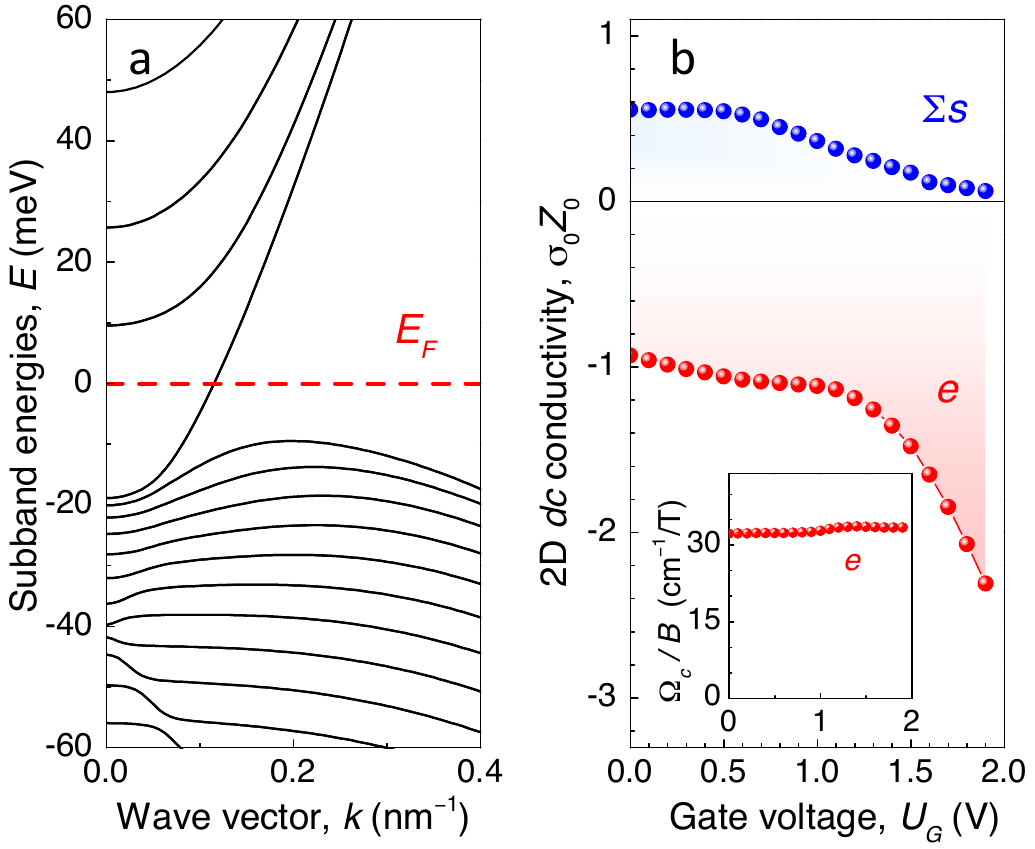}
\caption{\textbf{Charge carriers in strained HgTe.} \textbf{a}, The band structure of the Cd$_{0.7}$Hg$_{0.3}$Te/HgTe heterostructure close to the $\Gamma$-point. The chemical potential (dashed line) crosses the Dirac-like surface state in the band gap corresponding to the electron CR $\Omega_{ce}$. 
\textbf{b}, 2D $dc$ conductivity $\sigma_0$ of different charge carriers ($e$ and $\Sigma s = s1 + s2$), obtained by Drude-like fits to Eqs.~(\ref{sxx},\ref{sxy}) of the magnetooptical spectra. The dimensionless values are given relative to the impedance of free space $Z_0 = 1 /c  \epsilon_0 \approx 377 \, \mathrm{Ohm}$. 
The inset shows the $e$-CR in terms of $\Omega_{ce} / B$ as a function of the gate voltage $U_G$. }\label{fig2}
\end{figure}

Magnetic field dependence of the THz transmission is dominated by a sharp CR of surface electrons ($e$) $\Omega_{ce}$ at $B_{e} = 0.4 \, \mathrm{T}$ (Figs.~\ref{fig1}b and  \ref{fig1}c). Below we demonstrate their Dirac-like character and that they are responsible for the universal Faraday rotation. Remarkably, the observation of the CR both in $t_p$ and $t_c$  indicates a high purity of our HgTe layer. The scattering time is significantly longer than the inverse THz frequency $\omega \tau \gg 1$,  and according to Eqs.~(\ref{sxx}) and (\ref{sxy})  the $ac$ conductivity reveals a resonance-like behavior $\sigma_{xx}, \sigma_{xy} \propto 1/(\Omega_{ce}^2  - \omega^2)$. 

Further features are broad resonances  at $B_{s1} = 3.7 \, \mathrm{T}$ and at $B_{s2} = 2.2 \, \mathrm{T}$ indicated in Fig.~\ref{fig1} as $s1$ and $s2$, respectively. The  phase of the corresponding THz transmission coefficient $\phi_c$ in the vicinity of these resonances has the opposite sign with respect to that of the $e$-CR. 
Remarkably, the $s1$ and $s2$ resonances disappear with applying positive gate voltage (Fig.~\ref{fig1} and Fig.~\ref{fig2}b). We associate them with either interband Landau level transitions or thermally activated states as discussed below. 

To understand the origin of the experimentally observed resonances, we analyze the band structure of tensile strained $\text{Cd}_{0.7}\text{Hg}_{0.3}\text{Te / HgTe}$ layer as shown in Fig.~\ref{fig2}a.
It is obtained similar to Ref.~[\onlinecite{PhysRevX.4.041045}] within the tight binding approximation of the $6\times 6$ - Kane Hamiltonian \cite{novik_2005,baum_2014}. 
Due to reduced point symmetry at the boundary between the  $\text{Cd}_{0.7}\text{Hg}_{0.3}\text{Te}$ and $\text{HgTe}$ layers, an additional interface potential is allowed in the Hamiltonian   \cite{Roessler96}.  This potential is used to shift the Dirac point closer to the valence band edge, so that the tight binding results are in good agreement with recent ARPES experiments \cite{PhysRevLett.106.126803,Liu15} and ab-initio calculations \cite{Felser14} on HgTe.  
The Dirac-like surface states are located in the band gap between the light-hole (conduction) and heavy hole (valence) subbands (Fig.~\ref{fig2}a). The camel back of the heavy hole band originates from coupling of this band to the electron-like valence band and is therefore a hallmark of the inverted band structure of HgTe.
In accordance with previous transport data, the chemical potential crosses the topological surface states for a large range of gate voltages \cite{PhysRevX.4.041045}. The total electron density in Fig.~\ref{fig2}a is $n\approx 2 \times 10^{11} \, \mathrm{cm^{-2}}$, representing the experimental situation at  $U_G=1.9\,\mathrm{V}$. For simplicity, we assume here the same density at the top and bottom surfaces. Using the general formula for a classical cyclotron resonance  \cite{ashcroft_1976}  $\Omega_{c}=\frac{2\pi eB}{\hbar^{2}}\frac{\partial E\left(k\right)}{\partial A}$, where $E(k)$ is the energy dispersion, $B$ is the magnetic field, 
and  $A$ is the area enclosed by the wave vector $k$, we calculate for the topological surface state $\Omega_{ce} / B \approx 35 \,\mathrm{cm^{-1} / T}$. 

\begin{figure}[t]
\includegraphics[width=.46\textwidth]{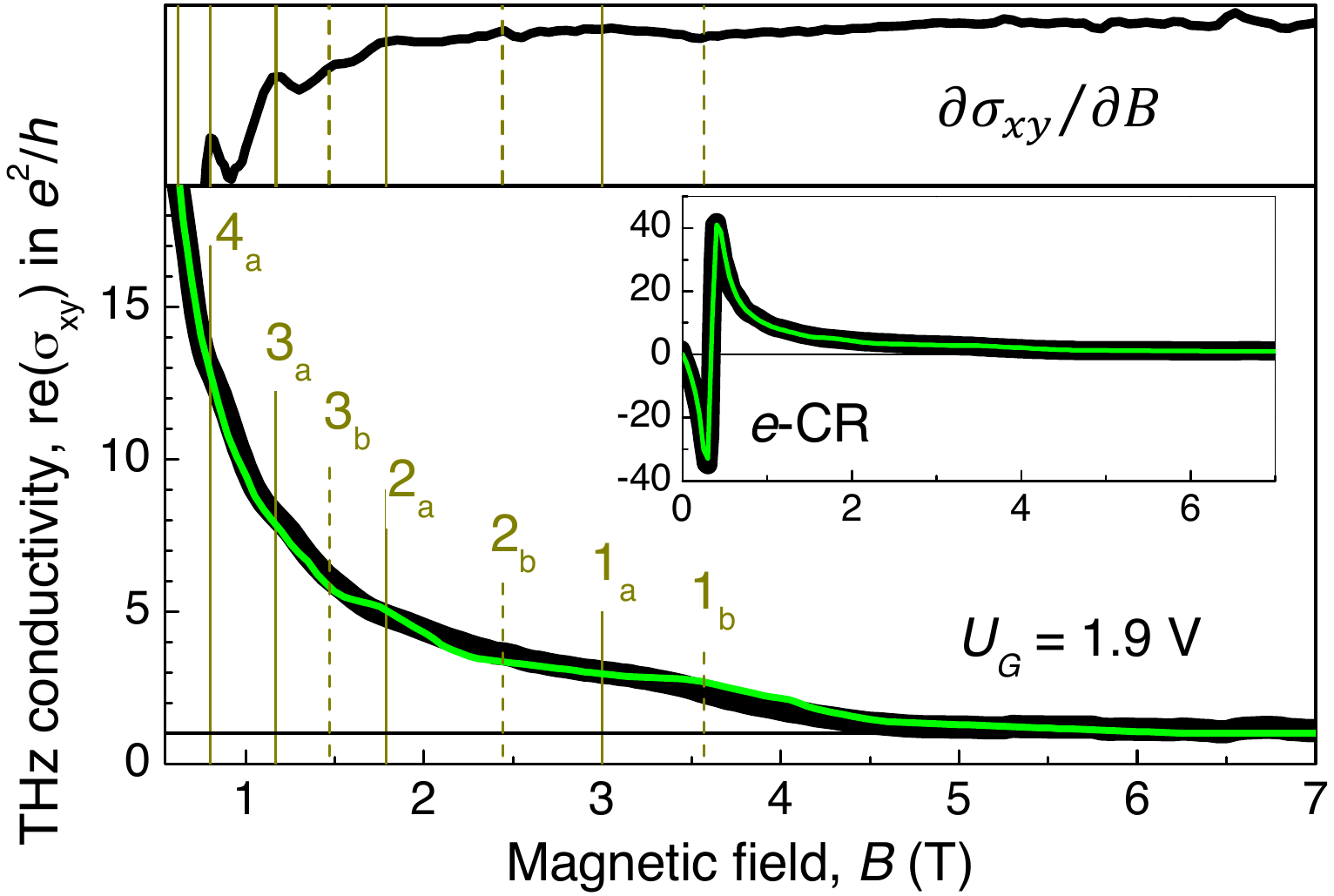}
\caption{\textbf{THz QHE of the surface states.} The real part of the THz Hall conductivity $\sigma_{xy}$ in units of $e^2/h$, obtained at $U_G = 1.9 \, \mathrm{V}$. The vertical solid and dashed lines indicate the positions of the Hall plateaus in two surfaces ($N_a$ and $N_b$), estimated from the extrema in $\partial \sigma_{xy} / \partial B$ (upper panel). Theoretical calculations represented by the thin line are performed as explained in the text. Inset presents the same experimental and theoretical curves in the whole magnetic field range, including the surface carrier CR at $0.4 \,  \mathrm{T}$. } \label{fig3}
\end{figure}

Experimentally, simultaneous fit of the real and imaginary parts of $t_{p}$ and $t_{c}$ allows the extraction of all transport characteristics, i.e., conductivity, charge carrier density, scattering time and CR frequency \cite{PhysRevLett.106.107404}. 
The inset of Fig.~\ref{fig2}b shows experimentally determined electron CR as a function of gate voltage, which perfectly agrees with the theoretical value for the topological Dirac-like surface states. Since only surface states are observed in transport experiments on the similar structures \cite{PhysRevX.4.041045}, a possible explanation of the appearance of additional resonances is interband Landau level transitions between heavy hole-like (HH) bulk bands and topological surface states. Such transitions are generally allowed as can be shown using the Kubo formula. Another possibility would be thermally activated transport between the camel back of the HH bulk band and the surface states. This is generally possible since the THz field may well induce heating of the carriers, resulting in a higher effective temperature compared to that of the lattice. The heating of the system would be consistent with the effective temperature of the surface states carriers  $T = 25 \, \mathrm{K}$,  as shown later in Fig.~\ref{fig3}.

From the obtained scattering time and the CR positions in the magnetooptical spectra of Figs.~\ref{fig1}b and  \ref{fig1}c,  one can calculate the mobility $\mu = \tau \Omega_c / B$.
The surface states demonstrate high mobility $\mu_e = 1.8 \times 10^{5} \, \mathrm{cm^{2} V^{-1} s^{-1}}$, which agrees with the $dc$ transport data. 
Since the $e$-CR and $s1$,$s2$-resonances occur at different magnetic fields, their contributions to the $ac$ transport can be clearly separated, as presented in Fig.~\ref{fig2}b. The striking feature of this plot is that the $ac$ conductivity of the surface states dominates at large gate voltages. 
In what follows, we concentrate therefore on $U_{G} > 1.0 \, \mathrm{V}$, while remaining weak contribution from the interband Landau-level transitions/thermally activated transitions are subtracted as explained in Methods.

Figure~\ref{fig3} demonstrates the real part of the Hall conductivity $\sigma_{xy}$. The overall behavior is provided by the high-field wing of the classical Drude model, i.e., Eq.~(\ref{sxy}), resulting in a rapid suppression of $\sigma_{xy}$ with growing magnetic field. In addition to the classical behaviour, regular oscillations in $\partial \sigma_{xy} / \partial B$ can be recognized, which are linear in inverse magnetic field. The slope of the linear behavior changes with gate voltage, reflecting gate dependence of the electron density in one of the two surfaces \cite{PhysRevB.87.121104}.  These QHE oscillations extrapolate to $N_a = 1/2$ value for large magnetic fields, demonstrating Dirac character of the surface electrons \cite{Buttner2011}. While the oscillations of $\partial \sigma_{xy} / \partial B$ in Fig.~\ref{fig3} are not clearly resolved, the visibility can be significantly improved by inserting the sample in a Fabry-P\'{e}rot resonator, as we have previously demonstrated for a similar structure \cite{PhysRevB.87.121104}. 

In magnetic fields above $5 \, \mathrm{T}$, the Hall conductivity clearly shows a plateau close to $\sigma_{xy} = e^2/h$, corresponding to  a value $(1/2) e^2/h$ per surface (Fig.~\ref{fig3}). Another plateau close to $3 e^2/h$ is also recognizable at a magnetic field of $3 \, \mathrm{T}$. The steps in $\sigma_{xy} $ loose their regularity in lower magnetic fields, as can be qualitatively explained by a finite THz frequency $\omega$ in magnetooptical experiments. As mentioned above, the overall behaviour of $\sigma_{xy} (\omega)$ is provided by the classical curve of Eq.~(\ref{sxy}), and the real part of $\sigma_{xy}$ can be approximated as $\sigma_{xy} (\omega) \approx \sigma_0 {\Omega_c}/[{(\Omega_{ce}^2 - \omega^2)\tau}]$, which in the limit $\Omega_{ce} \gg \omega$ reduces to the expression $\sigma_{xy}=ne/B$, being a multiple of $e^2/h$. In low magnetic fields, the CR frequency $\Omega_{ce}$ becomes comparable to the THz frequency $\omega$, destroying the regularities in $\sigma_{xy} (\omega)$.

Since in strained HgTe the Fermi level lies in the bulk band gap (see Fig.~2a and Ref.~\onlinecite{PhysRevX.4.041045}), we attribute the observed THz QHE to the formation of the 2D Landau levels at the top and bottom surfaces of the HgTe layer (Fig.~\ref{fig1}a).  This interpretation is further substantiated by our theoretical analysis of the $ac$ quantum Hall conductivity $\sigma_{xy}(\omega)$ calculated from the Kubo formula for both top and bottom surface states within the Dirac model  \cite{PhysRevB.82.161104, PSSB:PSSB201248385}. 

Our two-surface Dirac model describes well the surface carrier CR (the inset of Fig.~\ref{fig3}). The lengths of the theoretical Hall plateaus in the high magnetic field region (Fig.~\ref{fig3}) correlate correctly with the positions of the extrema in the derivative $\partial \sigma_{xy} / \partial B$. However, the model predicts much sharper transitions between the QHE plateaus, as observed in the experiment. One of two possible explanations is the heating of the surface carriers by the THz field, resulting in a higher effective temperature compared to that of the lattice. Such a heating can occur due to inefficient energy relaxation in the electronic system through the emission of LO phonons at low temperatures \cite{Heating}. The best fit of our experimental data is obtained with $T = 25 \, \mathrm{K}$ (Fig.~\ref{fig3}). Another explanation is based on spatial fluctuations of the surface carrier densities, which are likely to occur in our samples due to their  large lateral sizes compared to the typical Hall bars used in the $dc$ measurements. The experimental data of Fig.~\ref{fig3} can alternatively be well fitted assuming cold carriers ($T = 1.7 \, \mathrm{K}$) with density fluctuations within 10\% relative to their nominal values (Fig.~\ref{fig3}). As the fits are nearly indistinguishable, we cannot quantitatively determine the contributions of both mechanisms leading to the smearing of the THz QHE plateaus. 

The stronger the field, the closer the Hall conductivity to the quantized values expected for a two-surface Dirac system
\begin{equation}
\sigma_{xy} = (N_a + N_b + 1)\frac{e^2}{h}, \qquad N_{a,b}={\rm Int}\left( \frac{n_{a,b}\Phi_0}{|B|}\right),
\label{sigma_xy}
\end{equation}
where $N_{a,b}$ are the integer numbers of the highest occupied Landau levels at the top and bottom surfaces, with $n_{a,b}$ being the corresponding carrier densities 
($\Phi_0=h/|e|$ is the magnetic flux quantum). Upon approaching the cyclotron resonance, the Hall conductivity deviates from the quantized values in Eq.~(\ref{sigma_xy}) 
due to the predominance of the intraband transitions between the Landau levels.  
From the fitting procedure, we extract the nominal carrier densities 
$n_a = 0.92 \times 10^{11} \, \mathrm{cm^{-2}}$ and $n_b = 1.07 \times 10^{11} \, \mathrm{cm^{-2}}$. The total surface carrier density $n_a + n_b$ agrees well with that obtained from Drude-like fits of magnetooptical spectra. 
Another fitting parameter is the classical (Drude) surface conductivity $\sigma_a = \sigma_b \approx 50 e^2 / h$. 
Its large value indicates high surface carrier mobility, insuring that the condition for the quantum Hall regime, 
$\sqrt{2} \Omega_B \tau_{a,b}=4R_0\sigma_{a,b} \sqrt{ |B| / n_{a,b}\Phi_0 } > 1$, 
is met for $B > 1$T. Here, $\sqrt{2}\hbar\Omega_B=v(2\hbar|eB|)^{1/2}$ is the characteristic Landau level spacing for a Dirac system, 
$\tau_{a,b}$ are the scattering times of the top and bottom carriers, 
and $R_0=h/(2e^2)$ is the resistance quantum. 

\begin{figure}[t]
\includegraphics[width=.48\textwidth]{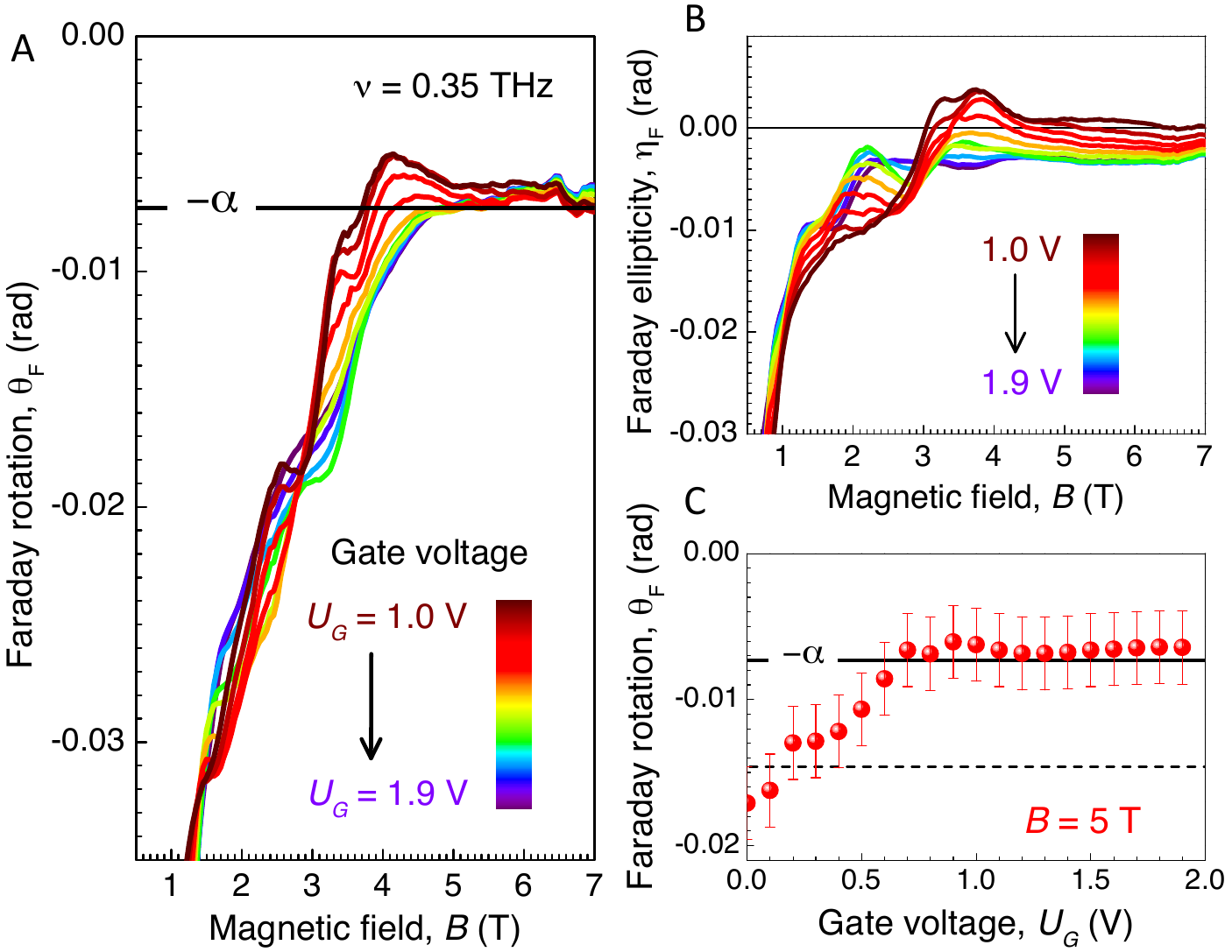}
\caption{\textbf{Quantized THz Faraday rotation of Dirac fermions.} \textbf{a}, Faraday rotation and \textbf{b}, Faraday ellipticity in a 3D HgTe TI as a function of the external magnetic field for different gate voltages (color-coded). The horizontal solid line in \textbf{a} indicates the universal Faraday rotation angle $\theta_F = - \alpha \approx - 7.3 \times 10^{-3} \, \mathrm{rad}$. \textbf{c}, Gate voltage dependence of the Faraday rotation in a magnetic field of $5 \, \mathrm{T}$.} \label{fig4}
\end{figure}

Having established that the THz response of the topological surface states in high magnetic fields $B > 5 \, \mathrm{T}$  is determined by the conductivity quantum $G_0 = e^2 / h$ ($N_a = N_b = 0$), we turn to the central result of this work, the THz Faraday effect.  Owing to the TME of Eq.~(\ref{Maxwell}), an oscillating electric $E_x e^{- i \omega t}$ (magnetic $H_y e^{- i \omega t}$) field of the linearly polarized THz radiation induces in a 3D TI  an oscillating magnetic $\alpha H_x e^{- i \omega t}$ (electric $\alpha E_y e^{- i \omega t}$) field. The generated in such a way secondary THz radiation is polarized perpendicular to the primary polarization and its amplitude is $\alpha$ times smaller. This can be viewed as a rotation of the initial polarization by an angle $| \theta_F | = \arctan \alpha \approx 7.3 \times 10^{-3} \, \mathrm{rad}$. Indeed,  Fig.~\ref{fig4}a clearly demonstrates that the Faraday angle in high magnetic fields is close to this fundamental value. 

We rigorously characterize the THz Faraday effect, and the Faraday ellipticity $\eta_F$ is shown in Fig.~\ref{fig4}b. It is relatively small $| \eta_F | < | \theta_F |$ in high magnetic fields, but does not reach zero. This observation indicates that while the TME dominates, the interaction of TIs with THz radiation is not a completely dissipationless process in our samples.  Remarkably, the universal value of the Faraday angle remains robust against the gate voltage. This is demonstrated in Fig.~\ref{fig4}c, where $| \theta_F | \approx \alpha$ for $0.7 \, \mathrm{V} < U_G < 1.9 \, \mathrm{V}$.  

The observed terahertz Faraday rotation equal to the fine structure constant $\alpha = 2 \pi e^2 / h c$ is a direct consequence of the topological magnetoelectric effect, confirming axion electrodynamics of 3D topological insulators. We use monochromatic terahertz spectroscopy, providing complete amplitude and phase reconstruction, which can be applied to investigate topological phenomena in various systems, including graphene, 2D electron gas, layered superconductors and recently experimentally discovered Weyl semimetals \cite{Weyl}. Picoradian angle resolution can be achieved using a balanced detection scheme \cite{PhysRevLett.104.036601}, and the universal Faraday rotation in combination with the magnetic flux quantum $\Phi_0=h/|e|$ and the conductivity quantum $G_0 = e^2 / h$ are suggested \cite{PhysRevLett.105.166803} to use for a metrological definition of the three basic physical constants, $e$, $h$, and $c$.

\section*{Acknowledgments}
This work was supported by Austrian Science Funds (I1648-N27, W-1243, P27098-N27), as well as by the SFB 1170  "ToCoTronics", the ENB Graduate School on Topological Insulators, the SPP 1666 and the ERC (project 3-TOP).

\appendix
 
\section*{Methods}
  
\subsection*{Theoretical analysis of magnetooptical spectra}

In general case, the light propagating along the $z$ direction can be
characterized by the orthogonal components of electric ($E_x$,
$E_y$) and magnetic ($H_x$, $H_y$) fields. These procedure closely
follows the formalism described by Berreman \cite{berreman_josa_1972}. We write the field components in the form of a 4D vector $\mathbf{V}$ 
\begin{equation} \mathbf{V} = \left(\begin{array}{c}
E_x \\
E_y \\
H_x \\
H_y
\end{array}\right) \,.
\end{equation}
The interconnection between vectors $\mathbf{V}_1$ and $\mathbf{V}_2$,
corresponding to different points in space separated by a distance
$\ell$, is given by $\mathbf{V}_2 = \hat{M}(\ell) \mathbf{V}_1$. Here, $\hat{M} (\ell)$
is a $4 \times 4$ transfer matrix. In case of isotropic dielectric substrate
the transfer matrix is given by 
\begin{widetext}
\begin{equation}
\hat{M}_{\mathrm{CdTe}}  (\ell) = \left(\begin{array}{cccc}
\cos(k\ell) & 0 & 0 & \imath Z \sin(k\ell) \\
0 & \cos(k\ell) & -\imath Z \sin(k\ell) & 0 \\
0 & -\imath Z^{-1} \sin(k\ell) & \cos(k\ell) & 0 \\
\imath Z^{-1} \sin(k\ell) & 0 & 0 & \cos(k\ell) \\
\end{array} \right) \,.
\label{eq2}
\end{equation}
\end{widetext}
Here, $Z=\sqrt{\mu / \varepsilon}$ and $k = \sqrt{\mu \varepsilon}
\, \omega / c$ is the wave vector. In the following $\mu=1$ is
assumed.

The particular choice of complex amplitudes as the tangential
components of electric and magnetic fields simplifies the treatment
of interfaces in case of normal incidence. The fields across the
interfaces are connected by the Maxwell equation $\nabla \times
\mathbf{H} =\hat{\sigma} \mathbf{E}$. Here, the $e^{-i\omega t}$ time dependence is
assumed for all fields and $\hat{\sigma}$ is the complex conductivity tensor of
the material. In the limit of thin film the transfer matrix reads
\begin{equation}
\hat{M}_{\mathrm{HgTe}}(\hat{\sigma}) = \left(\begin{array}{cccc}
1 & 0 & 0 & 0 \\
0 & 1 & 0 & 0 \\
Z_0 \sigma_{yx} & Z_0 \sigma_{yy} & 1 & 0 \\
-Z_0 \sigma_{xx} & -Z_0 \sigma_{xy} & 0 & 1 \\
\end{array} \right) \,.
\label{eq3}
\end{equation}
Here, $Z_0 = \sqrt{\mu_0/\varepsilon_0}
\approx 377$\ $\Omega$ is the impedance of free space. We note that
matrices $ \hat{M}_{\mathrm{CdTe}}$ and $\hat{M}_{\mathrm{HgTe}}$ as given in Eqs.~(\ref{eq2})-(\ref{eq3}) are fully equivalent
to well known transmission and reflection expressions for a bare
substrate and a thin film, respectively.

The total transfer matrix $\hat{M} =  \hat{M}_{\mathrm{CdTe}}  \hat{M}_{\mathrm{HgTe}}$
relates vectors $ \mathbf{V}$ in the air on both sides of the sample and
contains full information about transmission and reflection
coefficients. In order to calculate them, it is more convenient to change the basis.
In the new basis, the first component of the vector $ \mathbf{V}$ is the
amplitude of the linearly polarized wave ($E_x$) propagating in
positive direction, the second is that of the wave with the same
polarization propagating in negative direction, the third and the
fourth components are the same but for two waves with the orthogonal linear polarization
($E_y$). The propagation matrix in the new basis is $\hat{M}^{\prime} =
\hat{V}^{-1} \hat{M} \hat{V}$, with the basis transformation matrix being
 \begin{equation} \hat{V} = \left(\begin{array}{cccc}
1 & 1 & 0 & 0 \\
0 & 0 & 1 & 1 \\
0 & 0 & -1 & 1 \\
1 & -1 & 0 & 0 \\
\end{array} \right) \,. 
\end{equation}
The complex transmission ($t$) and reflection ($r$) coefficients for
a linearly polarized incident radiation could now be easily found
from the following equation
\begin{equation} \left(\begin{array}{c}
t_c \\
0 \\
t_p \\
0 \\
\end{array} \right) = \hat{M}^{\prime} \left(\begin{array}{c}
0 \\
r_c \\
1 \\
r_p \\
\end{array} \right). \label{eqtr} \end{equation}
Here, the $t_p$ and $t_c$ denote the complex transmittance amplitudes
within parallel and crossed polarizers, respectively. The same
conventions for reflectance are given by $r_p$ and $r_c$. 
Eq.~(\ref{eqtr}) can be inverted analytically to obtain the complex conductivity matrix from the transmission data. 

The Faraday rotation $\theta$ and Faraday ellipticity $\eta$ can be
directly obtained from the transmission amplitudes $|t_p|$, $|t_c|$ and
phases $\phi_p$, $\phi_c$ as
\begin{eqnarray}
\tan(2 \theta) &=& \, \frac{2 |t_p| |t_c| \cos(\phi_p - \phi_c)}
{|t_p|^2 - |t_c|^2},
\label{RotEll} \\
\sin(2 \eta) &=& \, \frac{2 |t_p| |t_c| \sin(\phi_p - \phi_c)}
{|t_p|^2 + |t_c|^2}. \nonumber
\end{eqnarray}

\subsection*{Subtraction of a background contribution}

We start from the directly measured spectra of Faraday rotation and ellipticity. We note, that related data on complex transmission $t_c$ and $t_p$ are known as well. Using the expression for the transmission of a film on a substrate, the Faraday rotation and ellipticity are directly inverted to obtain complex conductivities $\sigma_{xx}$ and $\sigma_{xy}$. In addition, from the fits of the transmission data, the parameters of all charge carriers are obtained. In the approximation of independent carriers, their conductivities are additive, and a weak background contribution associated with the $s$-resonances can be directly subtracted from $\sigma_{xx}$ and $\sigma_{xy}$. Finally, the corrected conductivity data are used to calculate the Faraday rotation and ellipticity from electron-like surface states.


\begin{thebibliography}{36}
\expandafter\ifx\csname natexlab\endcsname\relax\def\natexlab#1{#1}\fi
\expandafter\ifx\csname bibnamefont\endcsname\relax
  \def\bibnamefont#1{#1}\fi
\expandafter\ifx\csname bibfnamefont\endcsname\relax
  \def\bibfnamefont#1{#1}\fi
\expandafter\ifx\csname citenamefont\endcsname\relax
  \def\citenamefont#1{#1}\fi
\expandafter\ifx\csname url\endcsname\relax
  \def\url#1{\texttt{#1}}\fi
\expandafter\ifx\csname urlprefix\endcsname\relax\def\urlprefix{URL }\fi
\providecommand{\bibinfo}[2]{#2}
\providecommand{\eprint}[2][]{\url{#2}}

\bibitem[{\citenamefont{Kane and Mele}(2005)}]{PhysRevLett.95.226801}
\bibinfo{author}{\bibfnamefont{C.~L.} \bibnamefont{Kane}} \bibnamefont{and}
  \bibinfo{author}{\bibfnamefont{E.~J.} \bibnamefont{Mele}},
  \bibinfo{journal}{Phys. Rev. Lett.} \textbf{\bibinfo{volume}{95}},
  \bibinfo{pages}{226801} (\bibinfo{year}{2005}).

\bibitem[{\citenamefont{Bernevig et~al.}(2006)\citenamefont{Bernevig, Hughes,
  and Zhang}}]{Bernevig15122006}
\bibinfo{author}{\bibfnamefont{B.~A.} \bibnamefont{Bernevig}},
  \bibinfo{author}{\bibfnamefont{T.~L.} \bibnamefont{Hughes}},
  \bibnamefont{and} \bibinfo{author}{\bibfnamefont{S.-C.} \bibnamefont{Zhang}},
  \bibinfo{journal}{Science} \textbf{\bibinfo{volume}{314}},
  \bibinfo{pages}{1757} (\bibinfo{year}{2006}).

\bibitem[{\citenamefont{Fu and Kane}(2007)}]{PhysRevB.76.045302}
\bibinfo{author}{\bibfnamefont{L.}~\bibnamefont{Fu}} \bibnamefont{and}
  \bibinfo{author}{\bibfnamefont{C.~L.} \bibnamefont{Kane}},
  \bibinfo{journal}{Phys. Rev. B} \textbf{\bibinfo{volume}{76}},
  \bibinfo{pages}{045302} (\bibinfo{year}{2007}).

\bibitem[{\citenamefont{K\"onig et~al.}(2007)\citenamefont{K\"onig, Wiedmann,
  Br\"une, Roth, Buhmann, Molenkamp, Qi, and Zhang}}]{Koenig02112007}
\bibinfo{author}{\bibfnamefont{M.}~\bibnamefont{K\"onig}},
  \bibinfo{author}{\bibfnamefont{S.}~\bibnamefont{Wiedmann}},
  \bibinfo{author}{\bibfnamefont{C.}~\bibnamefont{Br\"une}},
  \bibinfo{author}{\bibfnamefont{A.}~\bibnamefont{Roth}},
  \bibinfo{author}{\bibfnamefont{H.}~\bibnamefont{Buhmann}},
  \bibinfo{author}{\bibfnamefont{L.~W.} \bibnamefont{Molenkamp}},
  \bibinfo{author}{\bibfnamefont{X.-L.} \bibnamefont{Qi}}, \bibnamefont{and}
  \bibinfo{author}{\bibfnamefont{S.-C.} \bibnamefont{Zhang}},
  \bibinfo{journal}{Science} \textbf{\bibinfo{volume}{318}},
  \bibinfo{pages}{766} (\bibinfo{year}{2007}).

\bibitem[{\citenamefont{Hsieh et~al.}(2008)\citenamefont{Hsieh, Qian, Wray,
  Xia, Hor, Cava, and Hasan}}]{Hsieh2008}
\bibinfo{author}{\bibfnamefont{D.}~\bibnamefont{Hsieh}},
  \bibinfo{author}{\bibfnamefont{D.}~\bibnamefont{Qian}},
  \bibinfo{author}{\bibfnamefont{L.}~\bibnamefont{Wray}},
  \bibinfo{author}{\bibfnamefont{Y.}~\bibnamefont{Xia}},
  \bibinfo{author}{\bibfnamefont{Y.~S.} \bibnamefont{Hor}},
  \bibinfo{author}{\bibfnamefont{R.~J.} \bibnamefont{Cava}}, \bibnamefont{and}
  \bibinfo{author}{\bibfnamefont{M.~Z.} \bibnamefont{Hasan}},
  \bibinfo{journal}{Nature} \textbf{\bibinfo{volume}{452}},
  \bibinfo{pages}{970} (\bibinfo{year}{2008}).

\bibitem[{\citenamefont{Qi et~al.}(2008)\citenamefont{Qi, Hughes, and
  Zhang}}]{PhysRevB.78.195424}
\bibinfo{author}{\bibfnamefont{X.-L.} \bibnamefont{Qi}},
  \bibinfo{author}{\bibfnamefont{T.~L.} \bibnamefont{Hughes}},
  \bibnamefont{and} \bibinfo{author}{\bibfnamefont{S.-C.} \bibnamefont{Zhang}},
  \bibinfo{journal}{Phys. Rev. B} \textbf{\bibinfo{volume}{78}},
  \bibinfo{pages}{195424} (\bibinfo{year}{2008}).

\bibitem[{\citenamefont{Tse and
  MacDonald}(2010{\natexlab{a}})}]{PhysRevLett.105.057401}
\bibinfo{author}{\bibfnamefont{W.-K.} \bibnamefont{Tse}} \bibnamefont{and}
  \bibinfo{author}{\bibfnamefont{A.~H.} \bibnamefont{MacDonald}},
  \bibinfo{journal}{Phys. Rev. Lett.} \textbf{\bibinfo{volume}{105}},
  \bibinfo{pages}{057401} (\bibinfo{year}{2010}{\natexlab{a}}).

\bibitem[{\citenamefont{Qi et~al.}(2009)\citenamefont{Qi, Li, Zang, and
  Zhang}}]{Qi27022009}
\bibinfo{author}{\bibfnamefont{X.-L.} \bibnamefont{Qi}},
  \bibinfo{author}{\bibfnamefont{R.}~\bibnamefont{Li}},
  \bibinfo{author}{\bibfnamefont{J.}~\bibnamefont{Zang}}, \bibnamefont{and}
  \bibinfo{author}{\bibfnamefont{S.-C.} \bibnamefont{Zhang}},
  \bibinfo{journal}{Science} \textbf{\bibinfo{volume}{323}},
  \bibinfo{pages}{1184} (\bibinfo{year}{2009}).

\bibitem[{\citenamefont{Hancock et~al.}(2011)\citenamefont{Hancock, van
  Mechelen, Kuzmenko, van~der Marel, Br\"une, Novik, Astakhov, Buhmann, and
  Molenkamp}}]{PhysRevLett.107.136803}
\bibinfo{author}{\bibfnamefont{J.~N.} \bibnamefont{Hancock}},
  \bibinfo{author}{\bibfnamefont{J.~L.~M.} \bibnamefont{van Mechelen}},
  \bibinfo{author}{\bibfnamefont{A.~B.} \bibnamefont{Kuzmenko}},
  \bibinfo{author}{\bibfnamefont{D.}~\bibnamefont{van~der Marel}},
  \bibinfo{author}{\bibfnamefont{C.}~\bibnamefont{Br\"une}},
  \bibinfo{author}{\bibfnamefont{E.~G.} \bibnamefont{Novik}},
  \bibinfo{author}{\bibfnamefont{G.~V.} \bibnamefont{Astakhov}},
  \bibinfo{author}{\bibfnamefont{H.}~\bibnamefont{Buhmann}}, \bibnamefont{and}
  \bibinfo{author}{\bibfnamefont{L.~W.} \bibnamefont{Molenkamp}},
  \bibinfo{journal}{Phys. Rev. Lett.} \textbf{\bibinfo{volume}{107}},
  \bibinfo{pages}{136803} (\bibinfo{year}{2011}).

\bibitem[{\citenamefont{Vald\'es~Aguilar
  et~al.}(2012)\citenamefont{Vald\'es~Aguilar, Stier, Liu, Bilbro, George,
  Bansal, Wu, Cerne, Markelz, Oh et~al.}}]{PhysRevLett.108.087403}
\bibinfo{author}{\bibfnamefont{R.}~\bibnamefont{Vald\'es~Aguilar}},
  \bibinfo{author}{\bibfnamefont{A.~V.} \bibnamefont{Stier}},
  \bibinfo{author}{\bibfnamefont{W.}~\bibnamefont{Liu}},
  \bibinfo{author}{\bibfnamefont{L.~S.} \bibnamefont{Bilbro}},
  \bibinfo{author}{\bibfnamefont{D.~K.} \bibnamefont{George}},
  \bibinfo{author}{\bibfnamefont{N.}~\bibnamefont{Bansal}},
  \bibinfo{author}{\bibfnamefont{L.}~\bibnamefont{Wu}},
  \bibinfo{author}{\bibfnamefont{J.}~\bibnamefont{Cerne}},
  \bibinfo{author}{\bibfnamefont{A.~G.} \bibnamefont{Markelz}},
  \bibinfo{author}{\bibfnamefont{S.}~\bibnamefont{Oh}}, \bibnamefont{et~al.},
  \bibinfo{journal}{Phys. Rev. Lett.} \textbf{\bibinfo{volume}{108}},
  \bibinfo{pages}{087403} (\bibinfo{year}{2012}).

\bibitem[{\citenamefont{Shuvaev
  et~al.}(2013{\natexlab{a}})\citenamefont{Shuvaev, Astakhov, Tkachov, Br\"une,
  Buhmann, Molenkamp, and Pimenov}}]{PhysRevB.87.121104}
\bibinfo{author}{\bibfnamefont{A.~M.} \bibnamefont{Shuvaev}},
  \bibinfo{author}{\bibfnamefont{G.~V.} \bibnamefont{Astakhov}},
  \bibinfo{author}{\bibfnamefont{G.}~\bibnamefont{Tkachov}},
  \bibinfo{author}{\bibfnamefont{C.}~\bibnamefont{Br\"une}},
  \bibinfo{author}{\bibfnamefont{H.}~\bibnamefont{Buhmann}},
  \bibinfo{author}{\bibfnamefont{L.~W.} \bibnamefont{Molenkamp}},
  \bibnamefont{and} \bibinfo{author}{\bibfnamefont{A.}~\bibnamefont{Pimenov}},
  \bibinfo{journal}{Phys. Rev. B} \textbf{\bibinfo{volume}{87}},
  \bibinfo{pages}{121104} (\bibinfo{year}{2013}{\natexlab{a}}).

\bibitem[{\citenamefont{Wu et~al.}(2013)\citenamefont{Wu, Brahlek,
  Valdes~Aguilar, Stier, Morris, Lubashevsky, Bilbro, Bansal, Oh, and
  Armitage}}]{Wu2013}
\bibinfo{author}{\bibfnamefont{L.}~\bibnamefont{Wu}},
  \bibinfo{author}{\bibfnamefont{M.}~\bibnamefont{Brahlek}},
  \bibinfo{author}{\bibfnamefont{R.}~\bibnamefont{Valdes~Aguilar}},
  \bibinfo{author}{\bibfnamefont{A.~V.} \bibnamefont{Stier}},
  \bibinfo{author}{\bibfnamefont{C.~M.} \bibnamefont{Morris}},
  \bibinfo{author}{\bibfnamefont{Y.}~\bibnamefont{Lubashevsky}},
  \bibinfo{author}{\bibfnamefont{L.~S.} \bibnamefont{Bilbro}},
  \bibinfo{author}{\bibfnamefont{N.}~\bibnamefont{Bansal}},
  \bibinfo{author}{\bibfnamefont{S.}~\bibnamefont{Oh}}, \bibnamefont{and}
  \bibinfo{author}{\bibfnamefont{N.~P.} \bibnamefont{Armitage}},
  \bibinfo{journal}{Nat. Phys.} \textbf{\bibinfo{volume}{9}},
  \bibinfo{pages}{410} (\bibinfo{year}{2013}).

\bibitem[{\citenamefont{Maciejko et~al.}(2010)\citenamefont{Maciejko, Qi, Drew,
  and Zhang}}]{PhysRevLett.105.166803}
\bibinfo{author}{\bibfnamefont{J.}~\bibnamefont{Maciejko}},
  \bibinfo{author}{\bibfnamefont{X.-L.} \bibnamefont{Qi}},
  \bibinfo{author}{\bibfnamefont{H.~D.} \bibnamefont{Drew}}, \bibnamefont{and}
  \bibinfo{author}{\bibfnamefont{S.-C.} \bibnamefont{Zhang}},
  \bibinfo{journal}{Phys. Rev. Lett.} \textbf{\bibinfo{volume}{105}},
  \bibinfo{pages}{166803} (\bibinfo{year}{2010}).

\bibitem[{\citenamefont{Tse and
  MacDonald}(2010{\natexlab{b}})}]{PhysRevB.82.161104}
\bibinfo{author}{\bibfnamefont{W.-K.} \bibnamefont{Tse}} \bibnamefont{and}
  \bibinfo{author}{\bibfnamefont{A.~H.} \bibnamefont{MacDonald}},
  \bibinfo{journal}{Phys. Rev. B} \textbf{\bibinfo{volume}{82}},
  \bibinfo{pages}{161104} (\bibinfo{year}{2010}{\natexlab{b}}).

\bibitem[{\citenamefont{Tkachov and Hankiewicz}(2011)}]{PhysRevB.84.035405}
\bibinfo{author}{\bibfnamefont{G.}~\bibnamefont{Tkachov}} \bibnamefont{and}
  \bibinfo{author}{\bibfnamefont{E.~M.} \bibnamefont{Hankiewicz}},
  \bibinfo{journal}{Phys. Rev. B} \textbf{\bibinfo{volume}{84}},
  \bibinfo{pages}{035405} (\bibinfo{year}{2011}).

\bibitem[{\citenamefont{Ikebe et~al.}(2010)\citenamefont{Ikebe, Morimoto,
  Masutomi, Okamoto, Aoki, and Shimano}}]{PhysRevLett.104.256802}
\bibinfo{author}{\bibfnamefont{Y.}~\bibnamefont{Ikebe}},
  \bibinfo{author}{\bibfnamefont{T.}~\bibnamefont{Morimoto}},
  \bibinfo{author}{\bibfnamefont{R.}~\bibnamefont{Masutomi}},
  \bibinfo{author}{\bibfnamefont{T.}~\bibnamefont{Okamoto}},
  \bibinfo{author}{\bibfnamefont{H.}~\bibnamefont{Aoki}}, \bibnamefont{and}
  \bibinfo{author}{\bibfnamefont{R.}~\bibnamefont{Shimano}},
  \bibinfo{journal}{Phys. Rev. Lett.} \textbf{\bibinfo{volume}{104}},
  \bibinfo{pages}{256802} (\bibinfo{year}{2010}).

\bibitem[{\citenamefont{Shimano et~al.}(2013)\citenamefont{Shimano, Yumoto,
  Yoo, Matsunaga, Tanabe, Hibino, Morimoto, and Aoki}}]{Shimano2013}
\bibinfo{author}{\bibfnamefont{R.}~\bibnamefont{Shimano}},
  \bibinfo{author}{\bibfnamefont{G.}~\bibnamefont{Yumoto}},
  \bibinfo{author}{\bibfnamefont{J.~Y.} \bibnamefont{Yoo}},
  \bibinfo{author}{\bibfnamefont{R.}~\bibnamefont{Matsunaga}},
  \bibinfo{author}{\bibfnamefont{S.}~\bibnamefont{Tanabe}},
  \bibinfo{author}{\bibfnamefont{H.}~\bibnamefont{Hibino}},
  \bibinfo{author}{\bibfnamefont{T.}~\bibnamefont{Morimoto}}, \bibnamefont{and}
  \bibinfo{author}{\bibfnamefont{H.}~\bibnamefont{Aoki}},
  \bibinfo{journal}{Nat. Commun.} \textbf{\bibinfo{volume}{4}},
  \bibinfo{pages}{1841} (\bibinfo{year}{2013}).

\bibitem[{\citenamefont{Br\"une et~al.}(2011)\citenamefont{Br\"une, Liu, Novik,
  Hankiewicz, Buhmann, Chen, Qi, Shen, Zhang, and
  Molenkamp}}]{PhysRevLett.106.126803}
\bibinfo{author}{\bibfnamefont{C.}~\bibnamefont{Br\"une}},
  \bibinfo{author}{\bibfnamefont{C.~X.} \bibnamefont{Liu}},
  \bibinfo{author}{\bibfnamefont{E.~G.} \bibnamefont{Novik}},
  \bibinfo{author}{\bibfnamefont{E.~M.} \bibnamefont{Hankiewicz}},
  \bibinfo{author}{\bibfnamefont{H.}~\bibnamefont{Buhmann}},
  \bibinfo{author}{\bibfnamefont{Y.~L.} \bibnamefont{Chen}},
  \bibinfo{author}{\bibfnamefont{X.~L.} \bibnamefont{Qi}},
  \bibinfo{author}{\bibfnamefont{Z.~X.} \bibnamefont{Shen}},
  \bibinfo{author}{\bibfnamefont{S.~C.} \bibnamefont{Zhang}}, \bibnamefont{and}
  \bibinfo{author}{\bibfnamefont{L.~W.} \bibnamefont{Molenkamp}},
  \bibinfo{journal}{Phys. Rev. Lett.} \textbf{\bibinfo{volume}{106}},
  \bibinfo{pages}{126803} (\bibinfo{year}{2011}).

\bibitem[{\citenamefont{Br\"une et~al.}(2014)\citenamefont{Br\"une, Thienel,
  Stuiber, B\"ottcher, Buhmann, Novik, Liu, Hankiewicz, and
  Molenkamp}}]{PhysRevX.4.041045}
\bibinfo{author}{\bibfnamefont{C.}~\bibnamefont{Br\"une}},
  \bibinfo{author}{\bibfnamefont{C.}~\bibnamefont{Thienel}},
  \bibinfo{author}{\bibfnamefont{M.}~\bibnamefont{Stuiber}},
  \bibinfo{author}{\bibfnamefont{J.}~\bibnamefont{B\"ottcher}},
  \bibinfo{author}{\bibfnamefont{H.}~\bibnamefont{Buhmann}},
  \bibinfo{author}{\bibfnamefont{E.~G.} \bibnamefont{Novik}},
  \bibinfo{author}{\bibfnamefont{C.-X.} \bibnamefont{Liu}},
  \bibinfo{author}{\bibfnamefont{E.~M.} \bibnamefont{Hankiewicz}},
  \bibnamefont{and} \bibinfo{author}{\bibfnamefont{L.~W.}
  \bibnamefont{Molenkamp}}, \bibinfo{journal}{Phys. Rev. X}
  \textbf{\bibinfo{volume}{4}}, \bibinfo{pages}{041045} (\bibinfo{year}{2014}).

\bibitem[{\citenamefont{Shuvaev et~al.}(2011)\citenamefont{Shuvaev, Astakhov,
  Pimenov, Br\"une, Buhmann, and Molenkamp}}]{PhysRevLett.106.107404}
\bibinfo{author}{\bibfnamefont{A.~M.} \bibnamefont{Shuvaev}},
  \bibinfo{author}{\bibfnamefont{G.~V.} \bibnamefont{Astakhov}},
  \bibinfo{author}{\bibfnamefont{A.}~\bibnamefont{Pimenov}},
  \bibinfo{author}{\bibfnamefont{C.}~\bibnamefont{Br\"une}},
  \bibinfo{author}{\bibfnamefont{H.}~\bibnamefont{Buhmann}}, \bibnamefont{and}
  \bibinfo{author}{\bibfnamefont{L.~W.} \bibnamefont{Molenkamp}},
  \bibinfo{journal}{Phys. Rev. Lett.} \textbf{\bibinfo{volume}{106}},
  \bibinfo{pages}{107404} (\bibinfo{year}{2011}).

\bibitem[{\citenamefont{Shuvaev
  et~al.}(2013{\natexlab{b}})\citenamefont{Shuvaev, Pimenov, Astakhov,
  MŸhlbauer, BrŸne, Buhmann, and Molenkamp}}]{APL4811496}
\bibinfo{author}{\bibfnamefont{A.}~\bibnamefont{Shuvaev}},
  \bibinfo{author}{\bibfnamefont{A.}~\bibnamefont{Pimenov}},
  \bibinfo{author}{\bibfnamefont{G.~V.} \bibnamefont{Astakhov}},
  \bibinfo{author}{\bibfnamefont{M.}~\bibnamefont{MŸhlbauer}},
  \bibinfo{author}{\bibfnamefont{C.}~\bibnamefont{BrŸne}},
  \bibinfo{author}{\bibfnamefont{H.}~\bibnamefont{Buhmann}}, \bibnamefont{and}
  \bibinfo{author}{\bibfnamefont{L.~W.} \bibnamefont{Molenkamp}},
  \bibinfo{journal}{Appl. Phys. Lett.} \textbf{\bibinfo{volume}{102}},
  \bibinfo{eid}{241902} (\bibinfo{year}{2013}{\natexlab{b}}).

\bibitem[{\citenamefont{Volkov et~al.}(1985)\citenamefont{Volkov, Goncharov,
  Kozlov, Lebedev, and Prokhorov}}]{volkov_infrared_1985}
\bibinfo{author}{\bibfnamefont{A.~A.} \bibnamefont{Volkov}},
  \bibinfo{author}{\bibfnamefont{Y.~G.} \bibnamefont{Goncharov}},
  \bibinfo{author}{\bibfnamefont{G.~V.} \bibnamefont{Kozlov}},
  \bibinfo{author}{\bibfnamefont{S.~P.} \bibnamefont{Lebedev}},
  \bibnamefont{and} \bibinfo{author}{\bibfnamefont{A.~M.}
  \bibnamefont{Prokhorov}}, \bibinfo{journal}{Infrared Phys.}
  \textbf{\bibinfo{volume}{25}}, \bibinfo{pages}{369} (\bibinfo{year}{1985}).

\bibitem[{\citenamefont{Shuvaev et~al.}(2012)\citenamefont{Shuvaev, Astakhov,
  Br\"{u}ne, Buhmann, Molenkamp, and Pimenov}}]{shuvaev_sst_2012}
\bibinfo{author}{\bibfnamefont{A.~M.} \bibnamefont{Shuvaev}},
  \bibinfo{author}{\bibfnamefont{G.~V.} \bibnamefont{Astakhov}},
  \bibinfo{author}{\bibfnamefont{C.}~\bibnamefont{Br\"{u}ne}},
  \bibinfo{author}{\bibfnamefont{H.}~\bibnamefont{Buhmann}},
  \bibinfo{author}{\bibfnamefont{L.~W.} \bibnamefont{Molenkamp}},
  \bibnamefont{and} \bibinfo{author}{\bibfnamefont{A.}~\bibnamefont{Pimenov}},
  \bibinfo{journal}{Semicond. Sci. Technol.} \textbf{\bibinfo{volume}{27}},
  \bibinfo{pages}{124004} (\bibinfo{year}{2012}).

\bibitem[{\citenamefont{Berreman}(1972)}]{berreman_josa_1972}
\bibinfo{author}{\bibfnamefont{D.~W.} \bibnamefont{Berreman}},
  \bibinfo{journal}{J. Opt. Soc. Am.} \textbf{\bibinfo{volume}{62}},
  \bibinfo{pages}{502} (\bibinfo{year}{1972}).

\bibitem[{\citenamefont{Palik and Furdyna}(1970)}]{palik_rpp_1970}
\bibinfo{author}{\bibfnamefont{E.~D.} \bibnamefont{Palik}} \bibnamefont{and}
  \bibinfo{author}{\bibfnamefont{J.~K.} \bibnamefont{Furdyna}},
  \bibinfo{journal}{Rep. Prog. Phys.} \textbf{\bibinfo{volume}{33}},
  \bibinfo{pages}{1193} (\bibinfo{year}{1970}).

\bibitem[{\citenamefont{Novik et~al.}(2005)\citenamefont{Novik,
  Pfeuffer-Jeschke, Jungwirth, Latussek, Becker, Landwehr, Buhmann, and
  Molenkamp}}]{novik_2005}
\bibinfo{author}{\bibfnamefont{E.~G.} \bibnamefont{Novik}},
  \bibinfo{author}{\bibfnamefont{A.}~\bibnamefont{Pfeuffer-Jeschke}},
  \bibinfo{author}{\bibfnamefont{T.}~\bibnamefont{Jungwirth}},
  \bibinfo{author}{\bibfnamefont{V.}~\bibnamefont{Latussek}},
  \bibinfo{author}{\bibfnamefont{C.~R.} \bibnamefont{Becker}},
  \bibinfo{author}{\bibfnamefont{G.}~\bibnamefont{Landwehr}},
  \bibinfo{author}{\bibfnamefont{H.}~\bibnamefont{Buhmann}}, \bibnamefont{and}
  \bibinfo{author}{\bibfnamefont{L.~W.} \bibnamefont{Molenkamp}},
  \bibinfo{journal}{Phys. Rev. B} \textbf{\bibinfo{volume}{72}},
  \bibinfo{pages}{035321} (\bibinfo{year}{2005}).

\bibitem[{\citenamefont{Baum et~al.}(2014)\citenamefont{Baum, B\"ottcher,
  Br\"une, Thienel, Molenkamp, Stern, and Hankiewicz}}]{baum_2014}
\bibinfo{author}{\bibfnamefont{Y.}~\bibnamefont{Baum}},
  \bibinfo{author}{\bibfnamefont{J.}~\bibnamefont{B\"ottcher}},
  \bibinfo{author}{\bibfnamefont{C.}~\bibnamefont{Br\"une}},
  \bibinfo{author}{\bibfnamefont{C.}~\bibnamefont{Thienel}},
  \bibinfo{author}{\bibfnamefont{L.~W.} \bibnamefont{Molenkamp}},
  \bibinfo{author}{\bibfnamefont{A.}~\bibnamefont{Stern}}, \bibnamefont{and}
  \bibinfo{author}{\bibfnamefont{E.~M.} \bibnamefont{Hankiewicz}},
  \bibinfo{journal}{Phys. Rev. B} \textbf{\bibinfo{volume}{89}},
  \bibinfo{pages}{245136} (\bibinfo{year}{2014}).

\bibitem[{\citenamefont{Ivchenko et~al.}(1996)\citenamefont{Ivchenko, Kaminski,
  and R\"ossler}}]{Roessler96}
\bibinfo{author}{\bibfnamefont{E.~L.} \bibnamefont{Ivchenko}},
  \bibinfo{author}{\bibfnamefont{A.~Y.} \bibnamefont{Kaminski}},
  \bibnamefont{and}
  \bibinfo{author}{\bibfnamefont{U.}~\bibnamefont{R\"ossler}},
  \bibinfo{journal}{Phys. Rev. B} \textbf{\bibinfo{volume}{54}},
  \bibinfo{pages}{5852} (\bibinfo{year}{1996}).

\bibitem[{\citenamefont{Liu et~al.}(2015)\citenamefont{Liu, Bian, Chang, Wang,
  Xu, Belopolski, Miotkowski, Cao, Miyamoto, Xu et~al.}}]{Liu15}
\bibinfo{author}{\bibfnamefont{C.}~\bibnamefont{Liu}},
  \bibinfo{author}{\bibfnamefont{G.}~\bibnamefont{Bian}},
  \bibinfo{author}{\bibfnamefont{T.-R.} \bibnamefont{Chang}},
  \bibinfo{author}{\bibfnamefont{K.}~\bibnamefont{Wang}},
  \bibinfo{author}{\bibfnamefont{S.-Y.} \bibnamefont{Xu}},
  \bibinfo{author}{\bibfnamefont{I.}~\bibnamefont{Belopolski}},
  \bibinfo{author}{\bibfnamefont{I.}~\bibnamefont{Miotkowski}},
  \bibinfo{author}{\bibfnamefont{H.}~\bibnamefont{Cao}},
  \bibinfo{author}{\bibfnamefont{K.}~\bibnamefont{Miyamoto}},
  \bibinfo{author}{\bibfnamefont{C.}~\bibnamefont{Xu}}, \bibnamefont{et~al.},
  \bibinfo{journal}{Phys. Rev. B} \textbf{\bibinfo{volume}{92}},
  \bibinfo{pages}{115436} (\bibinfo{year}{2015}).

\bibitem[{\citenamefont{Wu et~al.}(2014)\citenamefont{Wu, Yan, and
  Felser}}]{Felser14}
\bibinfo{author}{\bibfnamefont{S.-C.} \bibnamefont{Wu}},
  \bibinfo{author}{\bibfnamefont{B.}~\bibnamefont{Yan}}, \bibnamefont{and}
  \bibinfo{author}{\bibfnamefont{C.}~\bibnamefont{Felser}},
  \bibinfo{journal}{EPL (Europhysics Letters)} \textbf{\bibinfo{volume}{107}},
  \bibinfo{pages}{57006} (\bibinfo{year}{2014}).

\bibitem[{\citenamefont{Ashcroft and Mermin}(1976)}]{ashcroft_1976}
\bibinfo{author}{\bibfnamefont{N.~W.} \bibnamefont{Ashcroft}} \bibnamefont{and}
  \bibinfo{author}{\bibfnamefont{N.~D.} \bibnamefont{Mermin}},
  \emph{\bibinfo{title}{Solid State Physics}} (\bibinfo{publisher}{Harcourt
  College Publishers}, \bibinfo{year}{1976}).

\bibitem[{\citenamefont{Buttner et~al.}(2011)\citenamefont{Buttner, Liu,
  Tkachov, Novik, Brune, Buhmann, Hankiewicz, Recher, Trauzettel, Zhang
  et~al.}}]{Buttner2011}
\bibinfo{author}{\bibfnamefont{B.}~\bibnamefont{Buttner}},
  \bibinfo{author}{\bibfnamefont{C.~X.} \bibnamefont{Liu}},
  \bibinfo{author}{\bibfnamefont{G.}~\bibnamefont{Tkachov}},
  \bibinfo{author}{\bibfnamefont{E.~G.} \bibnamefont{Novik}},
  \bibinfo{author}{\bibfnamefont{C.}~\bibnamefont{Brune}},
  \bibinfo{author}{\bibfnamefont{H.}~\bibnamefont{Buhmann}},
  \bibinfo{author}{\bibfnamefont{E.~M.} \bibnamefont{Hankiewicz}},
  \bibinfo{author}{\bibfnamefont{P.}~\bibnamefont{Recher}},
  \bibinfo{author}{\bibfnamefont{B.}~\bibnamefont{Trauzettel}},
  \bibinfo{author}{\bibfnamefont{S.~C.} \bibnamefont{Zhang}},
  \bibnamefont{et~al.}, \bibinfo{journal}{Nat. Phys.}
  \textbf{\bibinfo{volume}{7}}, \bibinfo{pages}{418} (\bibinfo{year}{2011}).

\bibitem[{\citenamefont{Tkachov and Hankiewicz}(2013)}]{PSSB:PSSB201248385}
\bibinfo{author}{\bibfnamefont{G.}~\bibnamefont{Tkachov}} \bibnamefont{and}
  \bibinfo{author}{\bibfnamefont{E.~M.} \bibnamefont{Hankiewicz}},
  \bibinfo{journal}{physica status solidi (b)} \textbf{\bibinfo{volume}{250}},
  \bibinfo{pages}{215} (\bibinfo{year}{2013}).

\bibitem[{\citenamefont{Kiessling et~al.}(2012)\citenamefont{Kiessling, Quast,
  Kreisel, Henn, Ossau, and Molenkamp}}]{Heating}
\bibinfo{author}{\bibfnamefont{T.}~\bibnamefont{Kiessling}},
  \bibinfo{author}{\bibfnamefont{J.-H.} \bibnamefont{Quast}},
  \bibinfo{author}{\bibfnamefont{A.}~\bibnamefont{Kreisel}},
  \bibinfo{author}{\bibfnamefont{T.}~\bibnamefont{Henn}},
  \bibinfo{author}{\bibfnamefont{W.}~\bibnamefont{Ossau}}, \bibnamefont{and}
  \bibinfo{author}{\bibfnamefont{L.~W.} \bibnamefont{Molenkamp}},
  \bibinfo{journal}{Phys. Rev. B} \textbf{\bibinfo{volume}{86}},
  \bibinfo{pages}{161201} (\bibinfo{year}{2012}).

\bibitem[{\citenamefont{Xu et~al.}(2015)\citenamefont{Xu, Belopolski, Alidoust,
  Neupane, Bian, Zhang, Sankar, Chang, Yuan, Lee et~al.}}]{Weyl}
\bibinfo{author}{\bibfnamefont{S.-Y.} \bibnamefont{Xu}},
  \bibinfo{author}{\bibfnamefont{I.}~\bibnamefont{Belopolski}},
  \bibinfo{author}{\bibfnamefont{N.}~\bibnamefont{Alidoust}},
  \bibinfo{author}{\bibfnamefont{M.}~\bibnamefont{Neupane}},
  \bibinfo{author}{\bibfnamefont{G.}~\bibnamefont{Bian}},
  \bibinfo{author}{\bibfnamefont{C.}~\bibnamefont{Zhang}},
  \bibinfo{author}{\bibfnamefont{R.}~\bibnamefont{Sankar}},
  \bibinfo{author}{\bibfnamefont{G.}~\bibnamefont{Chang}},
  \bibinfo{author}{\bibfnamefont{Z.}~\bibnamefont{Yuan}},
  \bibinfo{author}{\bibfnamefont{C.-C.} \bibnamefont{Lee}},
  \bibnamefont{et~al.}, \bibinfo{journal}{Science}
  \textbf{\bibinfo{volume}{349}}, \bibinfo{pages}{613} (\bibinfo{year}{2015}).

\bibitem[{\citenamefont{Crooker et~al.}(2010)\citenamefont{Crooker, Brandt,
  Sandfort, Greilich, Yakovlev, Reuter, Wieck, and
  Bayer}}]{PhysRevLett.104.036601}
\bibinfo{author}{\bibfnamefont{S.~A.} \bibnamefont{Crooker}},
  \bibinfo{author}{\bibfnamefont{J.}~\bibnamefont{Brandt}},
  \bibinfo{author}{\bibfnamefont{C.}~\bibnamefont{Sandfort}},
  \bibinfo{author}{\bibfnamefont{A.}~\bibnamefont{Greilich}},
  \bibinfo{author}{\bibfnamefont{D.~R.} \bibnamefont{Yakovlev}},
  \bibinfo{author}{\bibfnamefont{D.}~\bibnamefont{Reuter}},
  \bibinfo{author}{\bibfnamefont{A.~D.} \bibnamefont{Wieck}}, \bibnamefont{and}
  \bibinfo{author}{\bibfnamefont{M.}~\bibnamefont{Bayer}},
  \bibinfo{journal}{Phys. Rev. Lett.} \textbf{\bibinfo{volume}{104}},
  \bibinfo{pages}{036601} (\bibinfo{year}{2010}).

\end{thebibliography}

\end{document}